\documentclass[12pt]{iopart}

\usepackage{iopams}
\newcommand{\be}{\begin{equation}}
\newcommand{\ee}{\end{equation}}
\newcommand{\beqn}{\begin{equation}}
\newcommand{\eeqn}{\end{equation}}
\newcommand{\bea}{\begin{eqnarray}}
\newcommand{\eea}{\end{eqnarray}}

\begin{document}

\title[Is chiral symmetry manifested in nuclear structure?]{Is chiral
symmetry manifested in nuclear structure?}

\author{R.J.\ Furnstahl$^{1}$ and A.\ Schwenk$^{2,3,4}$}

\address{$^{1}$ Department of Physics, The Ohio State University, 
         Columbus, OH 43210, USA}
\address{$^{2}$ TRIUMF, 4004 Wesbrook Mall, Vancouver, BC, V6T 2A3, Canada}
\address{$^{3}$ ExtreMe Matter Institute EMMI, GSI,
         64291 Darmstadt, Germany}
\address{$^{4}$ Institut f\"ur Kernphysik, TU Darmstadt,
         64289 Darmstadt, Germany}

\ead{furnstahl.1@osu.edu, schwenk@triumf.ca}

\begin{abstract}
Spontaneously broken chiral symmetry is an established property
of low-energy quantum chromodynamics, 
but finding direct evidence for it from nuclear structure
data is a difficult challenge.
Indeed, phenomenologically successful energy-density
functional approaches do not even have explicit pions.
Are there smoking guns for chiral symmetry in nuclei?
\end{abstract}

\section{Challenge}

Spontaneously broken chiral symmetry is a key property of 
quantum chromodynamics (QCD) at low energies.  It gives rise to pseudo-scalar
Goldstone bosons, the pions, which because of non-zero light quark
masses are not massless, but are much lighter than other hadrons. For
nuclear forces, the immediate consequence is the one-pion-exchange
(OPE) potential with its long range and characteristic tensor
structure. In more detail there are constraints on the form of
interactions in effective low-energy
Lagrangians~\cite{Bedaque:2002mn,Epelbaum:2005pn,Epelbaum:2008ga}. Since
nuclear structure is ultimately determined by low-energy QCD
and the nuclear Fermi momentum is comparable to $2m_\pi$, one
might expect stark evidence from data for the fingerprint of chiral
symmetry.  Thus we pose the question: Can we find smoking guns for
chiral symmetry in nuclei?

Because calculations based on chiral symmetry consequences (understood
broadly to range from including OPE to full consistency with chiral symmetry
constraints in low-energy Lagrangians) are becoming increasingly able
to describe nuclear structure, one might think the question is moot.
But we are not asking here whether theories or models consistent with
chiral symmetry are sufficient to describe data, but whether the
experimental data says this physics is necessary.  This is a more
difficult question and one with some counterexamples to the need for
explicit chiral symmetry.  At one extreme, pionless effective field
theory (EFT) describes few-body scattering at very low energies and
nuclear binding up to at least the alpha
particle~\cite{Bedaque:2002mn,Platter:2009pi}.  At the other extreme,
energy-density functional approaches to medium and heavy nuclei (such
as Skyrme) have wide phenomenological success without explicit
pions~\cite{Bender:2003jk}.

In trying to identify manifestations of chiral symmetry, it is
important to make the distinction between long-range chiral physics
and short-range physics, which may be free of chiral symmetry
constraints.  Nuclear structure will in general have contributions
from both.  To find a smoking gun, one will likely need to identify
observables (or combinations of observables) that are sensitive to the
long-range parts.

Because it is spontaneously broken, chiral symmetry most directly
relates processes with different numbers of pions.  Clear signatures
of chiral symmetry, even of the most basic pion properties (such as
OPE), can become a more subtle question for processes with no pions in
the initial or final state, which include nuclear structure properties
such as binding energies and spectra.  The signatures can be
particularly obscured for bulk properties, where the pion effects at
the mean-field level may be washed out by spin and isospin averaging.

\section{Smoking guns for chiral symmetry in other contexts}

For our challenge to find smoking guns in nuclear structure we adopt a
broad definition of evidence for chiral symmetry, which encompasses
the quantitative implications of a light pion and its quantum numbers
(such as OPE) as well as the more subtle effects that constrain
low-energy chiral Lagrangians.  Such smoking guns for spontaneously
broken chiral symmetry are readily found in data from other strong
interaction contexts.  The spectrum of mesons is the most obvious,
where $m_\pi^2 \ll m_\rho^2$ manifests the approximate Goldstone boson
nature of the pion and would-be parity doublets (e.g., the $\rho$ and
$a_1$) are split in mass. The physics of low-energy $\pi$--$\pi$ and
$\pi$--$N$ scattering, which is governed by chiral physics, provide
many other examples.  Here the successes of chiral perturbation theory
closely tie experiment and theory through chiral
symmetry~\cite{Gasser:1983yg,Bernard:2006gx}.

For nucleon-nucleon (NN) interactions, the situation is more
subtle. For example, the unnaturally large NN scattering lengths are
not related to chiral symmetry, but dominate the physics at very low
energies.  However, a compelling prototype for the sort of direct
evidence for chiral symmetry we seek from nuclear structure data can
be found in the partial-wave analyses of NN scattering by the Nijmegen
group and collaborators.  This example also lets us define more
concretely what we are looking for.  Simply an improved $\chi^2$ for a
fit to NN data using interactions with an input one-pion exchange
potential (or two-pion exchange) is not the strong and direct evidence
we prefer.  (After all, it is possible to fit this data accurately
with inverse scattering potentials with no pion exchange component.)
Rather it is the unbiased extraction through the fitting process of
pion parameters that we call a smoking gun.

As part of the analysis described in Ref.~\cite{Stoks:1992ja}, the masses of
the charged and neutral pions were used as fit parameters.  The
extracted values agreed with experiment within estimated one percent
errors.  At the same time the pion-nucleon coupling was obtained from
the fit and found consistent with extracted values from $\pi$--$N$
scattering~\cite{Stoks:1992ja,vanKolck:1997fu}.  The data provides
very clear signals for this coupling; indeed, the $pp\pi^0$ coupling
is said to be determined from each individual partial wave except for
$^1$S$_0$~\cite{Stoks:1992ja}.

Subsequent studies looked at the partial wave analysis with a finer
microscope and found direct evidence of two-pion exchange (TPE)
physics~\cite{Rentmeester:1999vw,Rentmeester:2003mf}.  Fitting the
pion mass again but now as a parameter in the TPE potential gave
agreement with experiment at the ten percent level.  Other evidence
includes consistency of the $c_i$ couplings of the sub-leading TPE
with determinations from $\pi$--$N$ scattering.  It is worth noting
that the sub-leading TPE provides the long-range part of three-nucleon
(3N) forces.  Finding smoking guns for chiral symmetry in nuclear
structure may therefore be tied to including 3N interactions.

In the example of NN phase shifts, the sensitivity to pion physics can
be enhanced by concentrating on the peripheral partial waves (e.g.,
G-waves and higher are fully explained by pion exchanges).  This
isolates the long-distance physics, which is chiral physics.  Even so,
a careful analysis of statistical and systematic errors together with
a large database was required to cleanly extract the direct evidence for
sub-leading chiral effects.  This highlights the difficulty in meeting
the challenge for nuclear structure, where the chiral origin
of mid-range physics from TPE may be difficult to resolve.

\section{Opportunities}

There are many opportunities to build on existing efforts in the
search for chiral smoking guns.  The growing applications of chiral EFT to
nuclear structure through no-core shell model, coupled cluster, and
lattice EFT calculations are building foundations for the desired
evidence from data and for investigations similar to the NN
partial-wave analyses. We emphasize again that while phenomenological
success starting from microscopic theories incorporating chiral
symmetry is gratifying, this type of indirect evidence is not what we
are looking for.  As the accuracy improves, however, we can ask
whether distinctions that point to pions and then more subtle chiral
symmetry constraints will become apparent (as with TPE in the
prototype partial-wave analysis).  For nuclear structure evidence,
many-body force contributions from long-range TPE may be key.

Other possibilities for smoking guns are in shell-model calculations
that start from interactions including pion physics, with monopole (or
more) adjustments fit to reproduce spectra and binding energies in a
major shell. Do the fits of shell-model interactions favor pion
exchanges for the long-range parts? If so, is it possible to isolate
data that are most sensitive to the long-range parts (e.g., to
constrain their sub-leading contributions)?

A third category may provide the greatest challenge to finding
explicit chiral symmetry: energy-density functionals (EDF).  Accurate
calculations of binding energies and other properties are made across
the mass table by EDF's such as Skyrme or Gogny, which do not have
explicit pions.  Relativistic EDF's in principle start with pions but
they do not contribute at the mean-field level used in relativistic
phenomenology.  Naive dimensional analysis for fit parameters in EDF's
(Skyrme or covariant) is suggestive of chiral
signatures~\cite{Friar:1995dt,Furnstahl:2001un} but is not
quantitative enough to be conclusive. We expect generalized EDF's will
be needed to find smoking guns.

There have been many studies that hint at a special role for chiral
symmetry in nuclear structure.  These include a wide array of models
that implement a partial restoration of chiral symmetry at finite
density, in-medium QCD sum rules, and Brown-Rho scaling (see for example
Refs.~\cite{Mosel:2008tz,Birse:1994cz}).  None of these rise as yet to
the level of a smoking gun for nuclear structure; the challenge is to
make them sufficiently quantitative and predictive based on data
(rather than model input).

Other direct possibilities for finding smoking guns of chiral symmetry
may lie with in-medium chiral EFT, which has been under recent
development (see Refs.~\cite{Epelbaum:2008ga,Weise:2007pf} and
references therein).  There are open questions, such as the
convergence of the perturbative approach, but it has the explicit
chiral ingredients needed.  We can ask: are there unique predictions from
long-range pion physics?  These may be difficult to find, as
illustrated by the subtleties in pinning down the origin of the
spin-orbit splittings in nuclei.  One might hope for the same type
fitting as in NN phase shifts.  A possibility already being considered
is the fitting of energy-density functionals that have been
supplemented with long-range chiral EFT physics.  This would mean
letting the pion parameters float in the fits in the same way as in
the Nijmegen phase shift analyses.  The answer is as yet unknown to
even the most basic question: Do medium and heavy nuclei know the pion
mass and coupling?

What can we say in general about where one should look for chiral
symmetry signatures?  Searching in bulk properties will be difficult
because of the averaging of pion contributions and the important role
of short-range physics.  The task is to identify observables
(analogous to peripheral partial waves) that isolate contributions
from long-range pion physics.  Past suggestions have included
single-particle energies, where the pion tensor force may be isolated;
isotope chains, where small differences from the isospin dependence
due to pion exchanges including long-range 3N forces may be amplified;
collective excitations with pionic quantum numbers; electroweak
axial-charge transitions and properties of pionic atoms (see for
example Refs.~\cite{Weise:2007pf,Brown:2001nh,Otsuka:2009cs}).

\section{Final comments}

Our challenge to nuclear structure theorists is to search for
\emph{direct} manifestations of chiral symmetry in nuclei.  Success in
this quest will help to unify descriptions of strong-interaction
phenomena, identify sensitivities that can suggest profitable new
experiments, guide microscopic calculations of nuclear properties
toward greater precision, and foreshadow what to expect under extreme
conditions.  It can also lead to ties to lattice QCD, whose growing
successes give rise to dreams of direct calculations of the lightest
nuclei, enabling new insights to chiral symmetry in nuclear structure.
However, the challenge to find smoking guns can also be constructively
answered in a negative way, by showing that explicit chiral symmetry
is \emph{not} necessary for particular nuclear structure observables.
So an alternative question to pose is: where are effective field
theories of nuclei without pionic degrees of freedom applicable?

\section*{Acknowledgments}

This work was supported in part by the National Science Foundation
under Grant No.~PHY--0653312, the UNEDF SciDAC Collaboration under
DOE Grant DE-FC02-07ER41457, the Natural Sciences and Engineering
Research Council of Canada (NSERC), and by the Helmholtz Alliance
Program of the Helmholtz Association, contract HA216/EMMI ``Extremes
of Density and Temperature: Cosmic Matter in the Laboratory''.  TRIUMF
receives federal funding via a contribution agreement through the
National Research Council of Canada.

\section*{References}

\bibliographystyle{unsrt}

\bibliography{vlowk_refs}

\end{document}